\documentclass[10pt,twoside]{article}
\usepackage[english]{babel}

\usepackage {graphicx}
\usepackage {graphics}
\usepackage {floatflt}
\usepackage {latexsym}
\usepackage {caption3} 
\usepackage[labelsep=period]{caption}[2007/11/04 v.3.1e]
\begin{document}

{\begin{center}
{\underline{\scriptsize{Journal of Siberian Federal University. Mathematics \& Physics 2011, 4(3), 275--281}}}
\end{center}}

\vspace{0.5cm}
{\raggedright
UDK 530.12:531.51}

\vspace{0.5cm}
{\raggedright
{\bf\Large The Demonstration of Gravitational Phase Transition inside of Fluid Static Ball \\}


\vspace{0.5cm}

{\raggedleft{\large \bfseries Alexandre~M.Baranov~\footnote{alex\_m\_bar@mail.ru; AMBaranov@sfu-kras.ru}}

{\raggedleft\small\mdseries Institute of the Engineering Physics
\& Radio-Electronics,\\
Siberian Federal University {\footnote{\copyright ~Siberian Federal University. All rights reserved}},\\
Svobodny Av. 79, Krasnoyarsk, 660041;\\
Siberian State Technology University,\\
Mira Av. 82, Krasnoyarsk, 660049, Russia\\}}}

{\begin{center}
{\underline{\scriptsize{Received 10.11.2010, received in revised form 10.12.2010, accepted 20.02.2011}}}
\end{center}}

{{\small{\it It is shown that at the center of gravitating fluid ball there is the phase transition of second kind of gravitational field as a change of field's algebraic type.

\bigskip
Keywords: a gravitational phase transition, Einstein's equations, General Relativity and Gravitation, algebraic classification of Petrov, 
the theory of catastrophes.}}}

\smallskip

\vspace{0.5cm}

\indent
The question about solutions of gravitational equations and researches of symmetric static models in General Relativity and Gravitation (GRG) today is also actual as yesterday. Petrov's algebraic classification of gravitational fields {\cite{pet1}} is known as one from high-power tools of gravitational field investigations. Therefore a consideration of this problem from more general positions allows to analyse the fluid ball behaviour from the point of view of the Petrov algebraic classification.

In this paper is considered a spherical model of a gravitating fluid ball from more general base than in the papers \cite{{bam1},{bamvzv1}}. Here we suppose that our model has spherical symmetry and is static, i.e. all metric functions don't depend from the time variable.

\section{Metric, Tetrads and Einstein equations}

A metric interval is chosen here in  Bondi's coordinates and written down as

\begin{equation} 
ds^2 = F(r)dt^2 +2L(r)dtdr-r^2(d{\theta}^2 + sin{\theta}^2d{\varphi}^2),
\label{eq:1}
\end{equation}

\noindent 
where functions $F=F(r)$ and $L=L(r)$ are the metric functions, $r$ is a radial variable; $\;t$ is a time variable; $\; \theta$ and $\varphi$ are the angle coordinates. The velocity of light  and Newton's gravitational constant $G_N$ are chosen equal to the unit. A determinant of a covariant metric tensor $g_{\alpha\beta}$, which corresponds to (\ref{eq:1}) equals to $det(g_{\alpha\beta})\equiv g=-L^{2}r^{4}sin^{2}\theta.$

We shall introduce the tetrads as four orthonormal basis vectors which base on the given metric (\ref{eq:1}): 

\begin{equation} 
g^{(0)}_{\mu}=\sqrt{F}\delta^{0}_{\mu}; \quad g_{(1)\mu}=L\delta^{1}_{\mu}+\frac{1}{2}F\;\delta^{0}_{\mu}; 
\label{eq:2}
\end{equation}

\begin{equation} 
g_{(2)\mu}=-\frac{r}{\sqrt{2}}(\delta^{2}_{\mu}+i\, sin\theta\;\delta^{3}_{\mu}); \quad g_{(3)\mu}=-\frac{r}{\sqrt{2}}(\delta^{2}_{\mu}-i\, sin\theta\;\delta^{3}_{\mu}); 
\label{eq:3}
\end{equation}

\begin{equation} 
g^{\mu}_{(0)}=L^{-1}\;\delta^{\mu}_{1}; \quad g^{\mu}_{(1)}=\delta^{\mu}_{0}+\frac{1}{2}FL^{-1}\;\delta^{\mu}_{1};
\label{eq:4}
\end{equation}

\begin{equation} 
g^{\mu}_{(2)}=\frac{1}{r\sqrt{2}}(\delta^{\mu}_{2}+\frac{i}{sin\theta}\;\delta^{\mu}_{3}); \quad g^{\mu}_{(3)}=\frac{1}{r\sqrt{2}}(\delta^{\mu}_{2}-\frac{i}{sin\theta}\;\delta^{\mu}_{3}),
\label{eq:5}
\end{equation}

\noindent
where $i$ is an imaginary unit and the greek indexes run over $0, 1, 2,3$. 

The Einstein equations in a tetrad form with a source in the form of the energy-momentum tensor (EMT) are written as

\begin{equation} 
G_{(\alpha)(\beta)}=R_{(\alpha)(\beta)}-\frac{1}{2}g_{(\alpha)(\beta)}R=-\kappa T_{(\alpha)(\beta)}, 
\label{eq:6}
\end{equation}
where $G_{(\alpha)(\beta)}$ are the tetrad components of the Einstein tensor; $\;R_{(\alpha)(\beta)}$ are the tetrad components of the Ricci tensor; $\;R =R^{(\alpha)}_{\quad(\alpha)}$ is the scalar curvature; $\kappa=8\pi$ is the Einstein constant in the select system of units. The EMT of the perfect pascal fluid is written down in this case as 

\begin{equation} 
T_{(\alpha)(\beta)} \equiv T_{(\alpha)(\beta)}^{fluid}=(\mu+p)u_{(\alpha)}u_{(\beta)}-pg_{(\alpha)(\beta)}\equiv\mu
u_{(\alpha)}u_{(\beta)}+pb_{(\alpha)(\beta)},
\label{eq:7}
\end{equation}

\noindent 
where $\mu(r)$ is a mass-energy density; $p(r)$ is a pressure of the perfect pascal fluid; 
$u_{(\alpha)}=g_{(\alpha)\mu} (dx^{\mu}/ds)$ is the 4-velocity in the tetrad form; $b_{(\alpha)(\beta)}=u_{(\alpha)}u_{(\beta)}-g_{(\alpha)(\beta)}$ is a 3-projector onto the spacelike 3-surface in the 4-dimensional world (3-metric); all functions depend from radial variable only. This 3-metric is orthogonal to the 4-velocity $b_{(\alpha)(\beta)}u^{(\alpha)}=0.$ 

We will rewrite Einstein's equations (\ref {eq:6}) in a form allowing to use properties of the energy-momentum tensor after using a connection between the scalar curvature and a trace of the energy-momentum tensor $R=\kappa T:$ 

\begin{equation} 
R_{(\alpha)(\beta)}=-\kappa\left(T_{(\alpha)(\beta)}-\frac{1}{2}g_{(\alpha)(\beta)}T\right).
\label{eq:8}
\end{equation}

This system of equations are rewritten as the system of three equations after a substitution of the metric functions into the system 

\begin{equation} 
\displaystyle\frac{\varepsilon}{x}(\ln{L})^{\prime}= \displaystyle\frac{\chi}{2}(\mu+p); 
\label{eq:9}
\end{equation}

\begin{equation} 
\displaystyle\frac{\varepsilon}{x}(\ln{L})^{\prime}-\frac{\varepsilon}{2}\left(\frac{F^{\prime\prime}}{F}+\frac{2}{x}(\ln{F})^{\prime}-{(\ln{F})^{\prime}}(\ln{L})^{\prime}\right)=-\chi p ; 
\label{eq:10}
\end{equation}

\begin{equation}
-\displaystyle\frac{1}{x^2}(1-\varepsilon)+\displaystyle\frac{\varepsilon}{x}\left(\ln{\frac{F}{L}}\right)^{\prime}=-\chi\frac{1}{2}(\mu -p),
\label{eq:11}
\end{equation}

\noindent 
where all derivatives are taken with respect of variable $x=r/R,$ $R$ is an exterior ball radius; $\chi = \kappa R^2.$ 

The functions: $\varepsilon(x),$  $F(x)$ and  $L(x)$ are connected as 

\begin{equation} 
\varepsilon(x)=\displaystyle\frac{F(x)}{L(x)^2}.
\label{eq:12}
\end{equation}

Such connection is appeared from an equality when the relational volume equals  zero  in a static case, $u^{\mu}_{\;\;;\mu} = 0.$ 

Now excluding the mass-energy density and the pressure from this system of equations, we get the linear differential equation with variable coefficients for function $G (x) $:

\begin{equation} 
G^{\prime \prime}+f(x)G^{\prime}+g(x)G=0,
\label{eq:13}
\end{equation}

\noindent
where $G=\sqrt{F},\;$ $f(x)=(\ln{\varphi})^{\prime}$, $\varphi(x)=\sqrt{\varepsilon}/x,$
and a coefficient $g(x)$ is equal to

\begin{equation} 
g(x)=\frac{2(1-\varepsilon)+x\varepsilon^{\prime}}{2x^2\varepsilon}.
\label{eq:14}
\end{equation}

If we will input a new variable $ \zeta = \zeta (x) $ according to a relation
\begin{equation} 
d{\zeta} = \displaystyle\frac{x dx}{\sqrt{\varepsilon(x)}},
\label{eq:15}
\end{equation}

\noindent
then the equation (\ref{eq:13}) goes over in the equation of a nonlinear spatial oscillator with a variable "frequency" $\Omega(\zeta(x))$ 

\begin{equation} 
G^{\prime \prime}_{\zeta \zeta}+\Omega^2(\zeta(x))G=0.
\label{eq:16}
\end{equation}
 
\noindent
As the equation (\ref {eq:16}) cannot be integrated, generally speaking, in elementary functions therefore the "frequency" $\;\Omega$ can be rewritten as derivative 
\begin{equation} 
\Omega^2 = - \displaystyle\frac{d}{dy}\left(\frac{\Phi}{y}\right), 
\label{eq:17}
\end{equation}

\noindent
where $y = x^2,\;$ and a function $\Phi$ is an analog of Newton's gravitational potential of an interior part of the fluid ball. The function $\Phi$ is found out of the gravitational equations through the function $\varepsilon$ as 

\begin{equation} 
\Phi=1-\varepsilon=\frac{\chi}{x}\int\mu(x)x^2 dx = \frac{\chi}{2\sqrt{y}}\int\mu(y)\sqrt{y} dy .
\label{eq:18}
\end{equation}

Besides we can easily find the relation for the pressure out of the system of gravitational equations (\ref{eq:9})-(\ref{eq:11})
\begin{equation} 
\chi p = -\displaystyle{\frac{\Phi}{x^2}+\frac{1}{x}}(1-\Phi)(\ln F)^{\prime}.
\label{eq:19}
\end{equation}

Now we will write down the function $\Omega^2$ as an expansion into series with respect to the variable $y$

\begin{equation} 
\Omega^2(y) = \sum_{n=0}^{\infty} a_n y^n,
\label{eq:20}
\end{equation}

here $y < 1.$

Further we will choose all $a_n $ equal to the zero, i.e. $ \Omega^2 = 0. $ Then (\ref {eq:16}) will be transformed into the equation 

\begin{equation} 
G^{\prime \prime}_{\zeta \zeta}=0
\label{eq:21}
\end{equation}

\noindent 
and function $G(\zeta) = C_1\cdot y+C_2,$ where $C_1, C_2$ are constants of an integration. 

In this case we have a homogeneous model of gravitational ball with a mass density $\mu_0 =const$ out of the relations (\ref{eq:16}) and (\ref{eq:17}), if $\Omega^2=0.$ The using of the connection between functions $\varepsilon,$ $F$ and $L$ (\ref {eq:12}) and also the condition of a joint with the exterior solution of Scharzschild \cite{syng1} leads to the well-known interior solution of Scharzschild \cite{syng1}.

Thereby we must remark that gravitational field of a spheric static distribution of the substance belongs to the algebraic type $D$ or the type $0$ of Petrov's classification \cite{pet1} in an accordance with the theorem \cite{bam2}.

In our case the gravitational field of the homogeneous distribution of a fluid has the algebraic type $0,$ i.e. the field is described by the conformally-flat solution of the Einstein equations.

We will put $n=0$ in the expansion (\ref{eq:20}) and now we have the constant "frequency" $\Omega^2 \equiv \Omega^2_0 = const.$ If $\Omega^2_0$ is positive ($\Omega^2_0 > 0$) then a general solution of equation (\ref{eq:15}) can be written down as a harmonic oscillating function

\begin{equation} 
G(\zeta(x)) = G_0 cos(\Omega_0 \zeta(x)+\varphi_0),
\label{eq:22}
\end{equation}

\noindent 
where $\varphi_0$ is a phase displacement.

We must here remark that the function $G(\zeta)$ describes spatial oscillations.

On the other hand we find the mass density out of (\ref{eq:17}) and (\ref{eq:18}) 

\begin{equation} 
\mu(x)=\mu_0 (1-b x^2),
\label{eq:23}
\end{equation}

\noindent 
which is the parabolic distribution, and function $G$ takes on the form 

\begin{equation} 
G(x) = G_0 cos(\Omega_0 (ln(\frac{d{\varepsilon(x)}}{dx})/(2 \sqrt{C_4} + \sqrt{\varepsilon(x)}))+\varphi_0),
\label{eq:24}
\end{equation}

\noindent 
where $\varepsilon(x) = 1 - C_3 x^2 +C_4 x^4; \;$ $C_3 = \chi R^2/3; \;$ $C_4 = \chi R^2 b/5.$

The using of gravitational equations with EMT of the perfect fluid for the given metric (\ref{eq:1}) simplifies a calculation of the Weyl tensor components. This tensor can be written in the tetrad components in general case as 

\begin{equation} 
W_{(\alpha)(\beta)(\gamma)(\delta)}= R_{(\alpha)(\beta)(\gamma)(\delta)} + R_{(\gamma)[(\alpha)}g_{(\beta)](\delta)} - R_{(\delta)[(\alpha)}g_{(\beta)](\gamma)} -\frac{1}{3} R g_{(\gamma)[(\alpha)} g_{(\beta)](\delta)},
\label{eq:25}
\end{equation}

\noindent
where the square brackets denote an antisymmetric operation with the indexes.

Mapping the Weyl tensor by 

\begin{equation} 
\Omega_{k}^{\alpha\beta} =\delta_{[{k}}^{\alpha}\delta_{{0}]}^{\beta} - \frac{i}{2} \varepsilon_{kmn} \delta_{m}^{\alpha} \delta_{n}^{\beta}
\label{eq:26}
\end{equation}

\noindent 
onto the 3-dimensional euclidean space with the metric $e_{ij} = diag(1,1,1),$ we will get the traceless $3 \times 3$ Weyl matrix (where $\varepsilon_{kmn}$ is an antisymmetric symbol Levi-Civita)

\begin{equation} 
W = \left(\frac{\eta}{2 R^2}\right) x^2 \times {diag(2,-1,-1)}.
\label{eq:27}
\end{equation}

\noindent
This matrix is a canonical form of Weyl's matrix of the algebraic type $D$ according to the algebraic classification of spaces for the researched case and for any $x \neq 0$ ($\eta = 2m/R $ is a compactness, $m$ is Schwarzschild's mass of the ball).

In the point $x = 0$ (at the center of ball) we have a degeneration of the Weyl matrix of D type into 0 type, i.e. a continual transition  $x \rightarrow 0$ be accompanied by a jumplike change of the algebraic type of space ($D \rightarrow 0.$) This change corresponds to the rank change of Weyl's matrix from $r = 3$ to $r = 0.$ Herewith the mass density of fluid is constant and equals $\mu_0$ into a neighborhood of point $x = 0.$ 
In other words the substance is not compressible nearby of the ball center and the algebraic type of gravitational field is the type $0$ (the conformally-flat space-time).

Thus the obtained result is in the full correspondence with the statement of the theorem \cite{bam2}. Moreover, the algebraic classification of gravitational fields is connected to the solution of a cubic characteristic  equation

\begin{equation} 
\lambda^3 + p \lambda + q = 0, 
\label{eq:28}
\end{equation}

\noindent
where the parameters $p$ and $q$ can be found for the given traceless matrix $W$ as

\begin{equation} 
p = - \frac {1}{2} Sp W^2; \qquad  q=-\frac{1}{3} Sp W^3.
\label{eq:29}
\end{equation}

The equation (\ref{eq:28}) can be considered as the extremum condition of some "potential" function

\begin{equation} 
V(\lambda, p, q) = V_0 + \frac{1}{4} {\lambda^4} +\frac{1}{2} p {\lambda^2} + q \lambda,
\label{eq:30}
\end{equation}

\noindent
where $V_0 = const.$

A deformation (\ref{eq:30}) describes the cusp catastrophe of Whitney in accordance with the catast-\linebreak
rophe theory. Solutions of the equation (\ref{eq:28}) are, generally speaking, in three areas, on three lines and in one point of plane of the control parameters $p$ and $q$ when we have the static case (see Fig.1).

\begin{figure}[t,h]
\center{\includegraphics[width=0.35\linewidth]{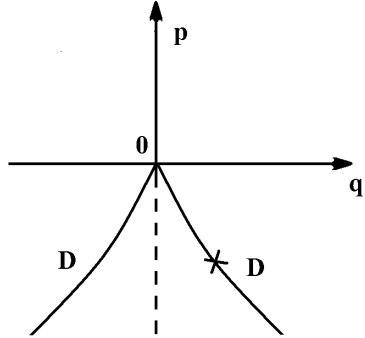}}
\caption{The cusp catastrophe's surface and its projection onto the plane of the control parameters $p$ and $q$.}
\end{figure}

It is known that the algebraic type of Riemann's space-time can be changed by the infinitesimal perturbations (herewith the rank of Weyl's matrix varies by jump) \cite{bam3}. The cusp ($p= q = 0$) accords with the phase transition into type 0 as into a most symmetric "phase" from point of view of the phase transitions of the second kind (Landau's phase theory \cite{dau1}). Furthermore the parameter $p$ is an analog of a temperature; a derivative  ${\partial{V}}/{\partial{p}}$ is the analog of the entropy; ${\partial^2{V}}/{\partial{p}}^2$ is the analog of the thermal capacity. Last derivative varies by jump in the cusp (when $p = 0$ ) also as the thermal capacity.

In our case the equation (\ref{eq:28}) has two real coincident roots, the sum of which equals the third root with an opposite sign:  $\lambda_2 = \lambda_3=\alpha_0 = (\eta/(2 R^2)) x^2;\;$ $\lambda_1 = -2 \alpha_0.$ The parameters $p$ and $q$ are expressed over through $\alpha_0$ as $p = 3 \alpha_0^2;\;$ $q = 2 \alpha_0^3.$ 
A discriminant $Q = (p/3)^3 + (q/2)^2 =0$ describes a semi-cubic parabola (a projection of the fold lines), which corresponds to space-time of $D$ type. The research case corresponds to a curve marked on Fig.1 by a cross because $q > 0$. 

Thus a state of system has not stability in the point of an inflection of function $V(\lambda, p, q)$ with $\lambda = \alpha_0$. 

The extremal values of function $V(\lambda, p, q)$ can be written down for all found roots of the equation (\ref{eq:28}) as

\begin{equation} 
V(\lambda_2, p) = V(\lambda_3, p) = \frac{1}{12} p^2 + C; \;\;\;\; V(\lambda_1, p) = -\frac{2}{3} p^2 + C,
\label{eq:31}
\end{equation}

\noindent
where $q$ and $\alpha_0$ are expressed through the parameter $p.$ There are the jumps of the second derivatives in respect of $p$ from the function $V$ in the point  $p = q =0:$ 
$\Delta (V_{,p,p}) = 1/6$ and  $\Delta (V_{,p,p}) = -4/3.$ These jumps correspond to the rank jump 
of the Weyl matrix in the cusp \cite{bam3}.

Hence the continual changes of the variable $x\;$ ($x \rightarrow 0$) and of parameters $p, q\;\;$ ($p \rightarrow 0,$ $q \rightarrow 0$) lead to the catastrophe: the algebraic type of space-time has the jump at the center of our ball model ($D \rightarrow 0$). 

From the viewpoint of the physical meaning there is the phase transition of second kind.
The eigenvalues $\lambda$ are the parameters of the order as in Landau's phase theory \cite{dau1}.

If we will rewrite the components of the Weyl matrix through the mass density then we can write down Weyl's matrix (\ref{eq:27}) by way of 

\begin{equation} 
W = \left(\frac{\eta}{2 R^2}\right)\left(1 -\frac{\mu(x)}{\mu_0} \right)\times {diag(2,-1,-1)}.
\label{eq:32}
\end{equation}

Here one can see the transition of matrix $W$ of the type $D$ into the matrix of type $0$ when the mass density $\mu(x)$ tends toward the constant value $\mu_0.$

The consideration of case with $n = 1$ leads to the expression $\;\Omega^2 =a_0 + a_1 y\;$ according to (\ref{eq:20}). Then the corresponding distribution of the mass density can be written as (see (\ref{eq:17})--(\ref{eq:18}))

\begin{equation} 
\mu(x)=\mu_0 (1-b x^2-c x^4),
\label{eq:33}
\end{equation}

\noindent
where $c$ is some constant.

The further analysis carried out similarly in a neighbourhood of the gravitating ball center leads to a previous result of the algebraic types' change by jump. The gravitational field has the algebraic type $0$ at the center (and into its small neighborhood) of the ball and in the remaining part of ball the gravitational field belongs to the type $D.$ In the expansion (\ref{eq:20}) there are not any new results for $n > 1.$ 

\section{Summary}

In the article the problem of an existence of another algebraic type at the fluid ball center than in the remaining part of ball is considered. The perfect fluid as the source of gravitational field has the arbitrary 
spherical distribution of the mass density. The eigenvalues task of Weyl's matrix is connected with the algebraic classification of Petrov. This classification is the algebraic classification of gravitational field in the Riemann space-time. At the center of ball the Weyl matrix of spherical distribution of the mass density equals to the zero, i.e. at the center the algebraic type is 0 (the conformally-flat space-time). 

On the other hand the theorem \cite{bam2} says that the spherical gravitational field can be only either the algebraic type $D$ or the type $0$ according to Petrov's algebraic classification. In general case the interior gravitational field of our fluid ball has the algebraic type $D$ outside of the center of ball, where we have 
the type $0.$

We can see from Fig.1 the cusp corresponds to the conformally-flat space-time (the algebraic type $0$). This correspondence is strong and the algebraic type is changed by jump there. It is the catastrophe or the phase transition of second kind in the physical meaning. The variables $\lambda_i$ (see (\ref{eq:28})) play here the role of the order parameter according to the phase theory of Landau \cite{dau1}.And the algebraic types of the Riemann space-time play the roles of  "the substance phases" (as in Landau's theory).

In other words into the small neighborhood of the ball center there is the phase transition of the second type in the spheric gravitational field of the fluid ball for any choice of $n\; (n \geq 0)$ in (\ref{eq:20}) as the  algebraic type change of Riemann's space-time.

\end{document}